\documentclass[prb,preprintnumbers,amsmath,amssymb,floatfix]{revtex4}

\usepackage{graphicx}
\usepackage{subfigure}
\usepackage{dcolumn}

\makeatletter

\begin{document}
\baselineskip 16pt

\title{Quantum Theory for Large Molecules $C_{60}$ Diffraction}
\author{Xiang-Yao Wu$^{a}$ \footnote{E-mail: wuxy2066@163.com},
Bai-Jun Zhang$^{a}$, Xiao-Jing Liu$^{a}$, Nuo Ba$^{a}$\\ Yi-Heng
Wu$^{a}$, Hou-Li Tang$^{a}$, Jing Wang$^{a}$ and Si-Qi Zhang$^{a}$
}

\affiliation{a.Institute of Physics, Jilin Normal University,
Siping 136000, China }

\begin{abstract}
Diffraction phenomena of large molecules have been studied in many
experiments, and these experiments are described by many
theoretical works. In this paper, we study $C_{60}$ molecules
single and double-slit diffraction with quantum theory approach,
and we pay close attention to the $C_{60}$ diffraction experiment
carried out by A.Zeilinger et.at in 1999. In double-slit
diffraction, we consider the decoherence effect, and find the
theoretical results
are good agreement with experimental data.\\
\vskip 5pt


\end{abstract}
\maketitle

\maketitle {\bf 1. Introduction} \vskip 8pt

As is well known, the matter-wave diffraction has become a large
field of interest over the last years, and it is extended to
electron, neutron, $C_{60}$, atom, more massive, complex objects,
like large molecules $I_{2}$, $C_{60}$ and $C_{70}$, which were
found in experiments [1-5]. At present, There are classical and
quantum methods to study interference and diffraction [6-12]. The
classical optics with its standard wave-theoretical methods and
approximations, in particular those of Huygens and Kirchhoff, has
been successfully applied to classical optics, and has yielded
good agreement with many experiments. This simple wave-optical
approach also gives a description of matter wave diffraction.
However, matter-wave interference and diffraction are quantum
phenomena, and its full description needs quantum mechanical
approach. Recently, there are some quantum theory approach to
study electron and $C_{60}$ diffraction, and obtain some important
and new results [13-17]. In this paper, we study the $C_{60}$
single and double-slit diffraction with the quantum approach, and
compare to the $C_{60}$ diffraction ($\lambda \approx 250$\AA)
carried out by A. Zeilinger \emph{et. at} in 1999 [18]. The
results we obtained are shown in Fig. 3, Fig. 4 and Fig. 5. In
view of quantum mechanics, the $C_{60}$ has the nature of wave,
and the wave is described by wave function $\psi(\vec{r},t)$, and
the wave function $\psi(\vec{r},t)$ has statistical meaning, i.e.,
$\mid\psi(\vec{r},t)\mid^{2}$ can be explained as particle's
probability density. For the $C_{60}$ slit diffraction, if we can
calculate the $C_{60}$ wave function $\psi(\vec{r},t)$
distributing on display screen, we can obtain the diffraction
intensity of $C_{60}$ molecules, since the diffraction intensity
is directly proportional to $\mid\psi(\vec{r},t)\mid^{2}$. In the
slit diffraction, the $C_{60}$ wave functions can be divided into
three parts. The first is the incident area, and the $C_{60}$ wave
function is a plane wave. The second is the slit area, where the
$C_{60}$ wave function can be calculated by the Schr\"{o}dinger
wave equation. The third is the diffraction area, where the
$C_{60}$ wave function can be obtained by path integral. In the
following, we shall calculate these wave functions. Finally, we
study the decoherence effect in the $C_{60}$ double slit
diffraction. The decoherence effect is important for the large
molecules diffraction, and the result is agreement with the
experiment data.
 \vskip 5pt
\newpage
 \setlength{\unitlength}{0.1in}
\begin{center}
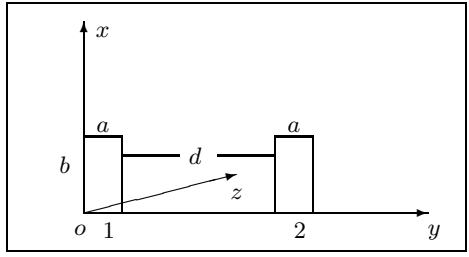
\begin{figure}
\begin{picture}(100,15)

 \put(26,4){\vector(1,0){18}}
 \put(26,4){\vector(0,1){10}}
 \put(26,4){\vector(4,1){8}}
 \put(22,2){\line(1,0){24}}
 \put(22,2){\line(0,1){13}}
 \put(46,2){\line(0,1){13}}
 \put(22,15){\line(1,0){24}}
 \put(26,8){\line(1,0){2}}
 \put(28,4){\line(0,1){4}}
 \put(36,4){\line(0,1){4}}
 \put(36,8){\line(1,0){2}}
 \put(38,4){\line(0,1){4}}
 \put(28,7){\line(1,0){3}}
 \put(33,7){\line(1,0){3}}
 \put(44,2.5){\makebox(2,1)[l]{$y$}}
 \put(26,13){\makebox(2,1)[c]{$x$}}
 \put(25.5,2.6){\makebox(2,1)[l]{$o$}}
 \put(33,4.5){\makebox(2,1)[c]{$z$}}
 \put(24,6){\makebox(2,1)[c]{$b$}}
 \put(26,8){\makebox(2,1)[c]{$a$}}
 \put(36,8){\makebox(2,1)[c]{$a$}}
 \put(31.5,6.5){\makebox(2,1)[l]{$d$}}
 \put(27,2.6){\makebox(2,1)[l]{$1$}}
 \put(37,2.6){\makebox(2,1)[l]{$2$}}
\end{picture}
\caption{Double-slit geometry with $a$ the  slit width,
 $b$ the slit length and $d$ the distance between the two slits.}
\label{moment}
\end{figure}
\end{center}
{\bf 2. Quantum approach of $C_{60}$ diffraction}

\vskip 8pt
 In an infinite plane, we consider a double-slit, its
width $a$, length $b$ and the slit-to-slit distance $d$ are shown
in FIG.\,1. The $x$ axis is along the slit length $b$ and the $y$
axis is along the slit width $a$. We calculate the $C_{60}$ wave
function in the first single slit (left) with the Schr\"odinger
equation, and the $C_{60}$ wave function of the second single slit
(right) can be obtained easily. At time $t$, we suppose that the
incident plane wave travels along the $z$ axis. It is
\begin{equation}
\psi_{0}(z, t)=Ae^{\frac{i}{\hbar}(pz-Et)},
\end{equation}
where $A$ is plane wave amplitude.

The potential in the first single slit is
\begin{eqnarray}
V(x,y,z)= \left \{ \begin{array}{ll}
   0  \hspace{0.3in} 0\leq x\leq b, 0\leq y\leq a, 0\leq z\leq c, \\
   \infty  \hspace{0.3in}  otherwise,
   \end{array}
   \right.
\end{eqnarray}
where $c$ is the thickness of the single slit. The time-dependent
and time-independent Schr\"odinger equations are
\begin{equation}
i\hbar\frac{\partial}{\partial
t}\psi(\vec{r},t)=-\frac{\hbar^{2}}{2M}(\frac{\partial^{2}}{\partial
x^{2}}+\frac{\partial^{2}}{\partial
y^{2}}+\frac{\partial^{2}}{\partial z^{2}})\psi(\vec{r},t),
\end{equation}
\begin{equation}
\frac{\partial^{2}\psi(\vec{r})}{\partial
x^{2}}+\frac{\partial^{2}\psi(\vec{r})}{\partial
y^{2}}+\frac{\partial^{2}\psi(\vec{r})}{\partial
z^{2}}+\frac{2ME}{\hbar^{2}}\psi(\vec{r})=0,
\end{equation}
where $M(E)$ is the mass(energy) of the $C_{60}$. The relation
between $\psi(\vec{r},t)$ and $\psi(\vec{r})$ is
\begin{equation}
\psi(\vec{r}_{1}, \vec{r}_{2}\cdots \vec{r}_{N},
t)=\psi(\vec{r}_{1}, \vec{r}_{2}\cdots \vec{r}_{N})g(t).
\end{equation}
The Eq. (19) become as

In Eq. (4), the wave function $\psi(x,y,z)$ satisfies the boundary
conditions
\begin{equation}
\psi(0,y,z)=\psi(b,y,z)=0,
\end{equation}
\begin{equation}
\psi(x,0,z)=\psi(x,a_{1},z)=0.
\end{equation}
The partial differential Eq. (4) can be solved by the method of
separation of variable. By writing

\begin{equation}
\psi(x,y,z)=X(x)Y(y)Z(z).
\end{equation}
The general solution of Eq. (3) is
\begin{eqnarray}
\psi_{1}(x,y,z,t)&=&\sum_{mn}\psi_{mn}(x,y,z,t) \nonumber\\
&=&\sum_{mn}D_{mn}\sin{\frac{n\pi x}{b}}\sin{\frac{m\pi
y}{a_{1}}}e^{i\sqrt{\frac{2ME}{\hbar^{2}}-\frac{n^{2}\pi^{2}}{b^{2}}-\frac{m^{2}\pi^{2}}{a_{1}^{2}}}z}e^{-\frac{i}{\hbar}Et}.
\end{eqnarray}
Equation (9) is the $C_{60}$ wave function in the first single
slit. Since the wave functions are continuous at $z=0$, we have
\begin{equation}
\psi_{0}(x,y,z,t)\mid_{z=0}=\psi_{1}(x,y,z,t)\mid_{z=0},
\end{equation}
from Eqs. (2), (6) and (9), we can obtain the Fourier coefficient
$D_{mn}$ by Fourier transform
\begin{eqnarray}
D_{mn}&=&\frac{4}{a_{1}
b}\int^{a_{1}}_{0}\int^{b}_{0}A\sin{\frac{n\pi
\xi}{b}}\sin{\frac{m\pi \eta}{a_{1}}}d\xi d\eta \nonumber\\
&=&\left \{ \begin{array}{ll}
   \frac{16A}{mn\pi^{2}} \hspace{0.6in} m,n, odd, \\
   0 \hspace{0.9in} otherwise,
   \end{array}
   \right.
\end{eqnarray}
substituting Eq. (11) into Eq. (9), we can obtain the $C_{60}$
wave function in the first single slit.

\begin{eqnarray}
\psi_{1}(x,y,z,t)&=&\sum_{m,n=0}^{\infty}\frac{16A}{(2m+1)(2n+1)\pi^{2}}\sin{\frac{(2n+1)\pi
x}{b}}\sin{\frac{(2m+1)\pi y}{a}} \nonumber\\&& \cdot
 e^{i\sqrt{\frac{2ME}{\hbar^{2}}-\frac{(2n+1)^{2}\pi^{2}}{b^{2}}
-\frac{(2m+1)^{2}\pi^{2}}{a^{2}}}z}e^{-\frac{i}{\hbar}Et}.
\end{eqnarray}

The $C_{60}$ wave function in the second single slit can be
obtained by making the coordinate translations $x'=x$, $y'=y-a-d$,
$z'=z$, and we can obtain the $C_{60}$ wave function
$\psi_{2}(x,y,z,t)$ in the second slit
\begin{eqnarray}
\psi_{2}(x,y,z,t)&=&\sum_{m,n=0}^{\infty}\frac{16A}{(2m+1)(2n+1)\pi^{2}}
\sin{\frac{(2n+1)\pi x}{b}}\sin{\frac{(2m+1)\pi (y-a_{1}-d)}{a}}
\nonumber\\&& \cdot
 e^{i\sqrt{\frac{2ME}{\hbar^{2}}-\frac{(2n+1)^{2}\pi^{2}}{b^{2}}
-\frac{(2m+1)^{2}\pi^{2}}{a^{2}}}z}e^{-\frac{i}{\hbar}Et}.
\end{eqnarray}
\vskip 8pt

{\bf 3. The wave function of $C_{60}$ diffraction}

\vskip 8pt With the approach of path integral, we can calculate
 $C_{60}$ wave function in the diffraction area.

The diffraction area is shown in FIG. 2, where $\mathbf{r}_{0}$ is
the position of point $P_{0}(x_{0}, y_{0}, c)$ on the slit surface
$(z=c)$, $P(x, y, z)$ is an arbitrary point in the diffraction
area. From Eq. (12), we can obtain the wave function in slit
surface $(z=c)$
\begin{eqnarray}
\psi_{P_{0}}(x,y,z,t)&=&\sum_{m,n=0}^{\infty}\frac{16A}{(2m+1)(2n+1)\pi^{2}}\sin{\frac{(2n+1)\pi
x_{0}}{b}}\sin{\frac{(2m+1)\pi y_{0}}{a}} \nonumber\\&& \cdot
 e^{i\sqrt{\frac{2ME}{\hbar^{2}}-\frac{(2n+1)^{2}\pi^{2}}{b^{2}}
-\frac{(2m+1)^{2}\pi^{2}}{a^{2}}}c}e^{-\frac{i}{\hbar}Et_{0}}.
\end{eqnarray}
The diffraction wave function $\psi_{out}$ on the point $P(x, y,
z)$ can be calculated by the formula of path integral [17]
\begin{equation}
\psi_{P}({\pmb r}, t)=\int K({\pmb r},t; {\pmb r_{0}},
t_{0})\psi_{P_{0}}({\pmb r_{0}}, t_{0})dx_{0}dy_{0},
\end{equation}
where $K({\pmb r},t; {\pmb r_{0}},t_{0})$ is the $C_{60}$
molecules propagater, it is
\begin{equation}
 K({\pmb r},t;
{\pmb r_{0}},t_{0})=(\frac{M}{2\pi i \hbar
(t-t_{0})})^{\frac{3}{2}}exp[\frac{iMR^{2}}{2\hbar (t-t_{0})}],
\end{equation}
\setlength{\unitlength}{0.1in}
 \begin{center}
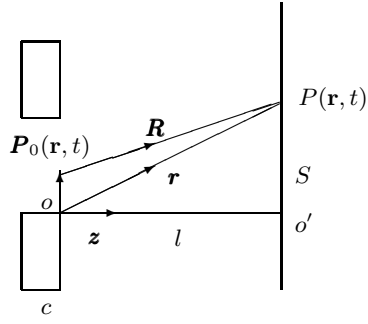
\begin{figure}
\begin{picture}(100,10)
 \put(26,5){\vector(1,0){3}}
 \put(26,5){\vector(0,1){2.2}}
 \put(26,5){\vector(2,1){5}}
 \put(24,1){\line(1,0){2}}
 \put(24,1){\line(0,1){4}}
 \put(26,1){\line(0,1){4}}
 \put(24,5){\line(1,0){2}}
 \put(24,10){\line(1,0){2}}
 \put(24,10){\line(0,1){4}}
 \put(26,10){\line(0,1){4}}
 \put(24,14){\line(1,0){2}}
 \put(37.6,1){\line(0,1){15}}
 \put(26,5){\line(1,0){11.6}}
 \put(26,7){\line(3,1){11.5}}
 \put(26,7){\vector(3,1){5}}
 \put(26,5){\line(2,1){11.5}}
 \put(27.5,3.2){\makebox(2,1)[l]{${\pmb z}$}}
 \put(32,3.2){\makebox(2,1)[l]{${ l}$}}
 \put(24.5,8){\makebox(2,1)[c]{${\pmb P_{0}(\mathbf{r},t)}$}}
 \put(25,5){\makebox(2,1)[l]{$o$}}
 \put(31,6.3){\makebox(2,1)[c]{${\pmb r}$}}
 \put(30,9){\makebox(2,1)[c]{${\pmb R}$}}
 \put(38.5,10.5){\makebox(2,1)[l]{$P(\mathbf{r},t)$}}
 \put(25,-0.5){\makebox(2,1)[l]{$c$}}
 \put(38.3,4){\makebox(2,1)[l]{${o'}$}}
 \put(38.3,6.5){\makebox(2,1)[l]{${S}$}}
\end{picture}
\caption{Diffraction area of the single slit} \label{moment}
\end{figure}
\end{center}
where $M$ is the mass of $C_{60}$ molecules, and $R$ is the
distance between point $P_{0}$ and point $P$, it is
\begin{eqnarray}
R&=&\sqrt{|\mathbf{r}-\mathbf{r}_{0}|^{2}}=\sqrt{(x-x_{0})^{2}+(y-y_{0})^{2}+(z-c)^{2}}
\nonumber\\&=&\sqrt{x^{2}+y^{2}+z^{2}+x_{0}^{2}+y_{0}^{2}+c^{2}-2xx_{0}-2yy_{0}-2zc}
.
\end{eqnarray}
Assume that the angle between $\mathbf{r}$ and $x$ axis ($y$ axis)
is $\frac{\pi}{2}-\alpha$ $(\frac{\pi}{2}-\beta)$, and then
$\alpha$ ($\beta$) is the angle between $\mathbf{r}$ and the
surface of $yz (xz)$. we obtain
\begin{eqnarray}
R^{2}&=&
x^{2}+y^{2}+z^{2}+x_{0}^{2}+y_{0}^{2}+c^{2}-2xx_{0}-2yy_{0}-2zc\nonumber\\&=&r^{2}-2r
\sin\alpha x_{0}-2r \sin\beta
y_{0}+x_{0}^{2}+y_{0}^{2}\nonumber\\&\approx&r^{2}-2r \sin\alpha
x_{0}-2r \sin\beta y_{0},
\end{eqnarray}
where $r^{2}=x^{2}+y^{2}+(z-c)^{2}$, In Eq. (18), the terms
$x_{0}^{2}$ and $y_{0}^{2}$ can be neglected. Substituting Eqs.
(14), (16), (18) into (15) yields
\begin{eqnarray}
\psi_{P}({\pmb r},t)&=&e^{\frac{iMr^{2}}{2\hbar(t-t_{0})}
}(\frac{M}{2\pi i\hbar
(t-t_{0})})^{\frac{3}{2}}\sum_{m=0}^{\infty}\sum_{n=0}^{\infty}\frac{16A}{(2m+1)(2n+1)\pi^{2}}
\nonumber\\&& \cdot
e^{i\sqrt{\frac{2ME}{\hbar^{2}}-(\frac{(2n+1)\pi}{b})^{2}-(\frac{(2m+1)\pi}{a_{1}})^{2}}\cdot
c}\nonumber\\&& \ \int^{b}_{0}e^{\frac{-iMr\sin\alpha
x_{0}}{\hbar(t-t_{0})}}\sin \frac{(2n+1)\pi x_{0}}{b}dx_{0}
\int^{a}_{0}e^{\frac{-iMr\sin\beta y_{0}}{\hbar(t-t_{0})}} \sin
\frac{(2m+1)\pi y_{0}}{a}dy_{0}.
\end{eqnarray}
With de Broglie relationship $p=\hbar k=\frac{h}{\lambda}$, there
is
\begin{eqnarray}
k=\frac{M v}{\hbar},
\end{eqnarray}
where $v=\frac{R}{t-t_{0}}$, thus
\begin{eqnarray}
k=\frac{M R}{\hbar (t-t_{0})}\approx \frac{M r}{\hbar (t-t_{0})},
\end{eqnarray}
then

\begin{eqnarray}
(\frac{M}{2\pi i\hbar (t-t_{0})})^{\frac{3}{2}}=(\frac{Mr}{2\pi
i\hbar (t-t_{0})
r})^{\frac{3}{2}}=(-\frac{\sqrt{2}}{2}-\frac{\sqrt{2}}{2}i)(\frac{k}{2
\pi r})^{\frac{3}{2}},
\end{eqnarray}
and
\begin{equation}
e^{\frac{iMr^{2}}{2\hbar (t-t_{0})}}\approx e^{\frac{ikr}{2}},
\end{equation}
and so
\begin{eqnarray}
\psi_{1 P}({\pmb
r},t)&=&(-\frac{\sqrt{2}}{2}-\frac{\sqrt{2}}{2}i)(\frac{k}{2 \pi
r})\sqrt{\frac{k}{2 \pi r}}e^{\frac{ikr}{2}}
\sum_{m=0}^{\infty}\sum_{n=0}^{\infty}\frac{16A}{(2m+1)(2n+1)\pi^{2}}
\nonumber\\&& \cdot
e^{i\sqrt{\frac{2ME}{\hbar^{2}}-(\frac{(2n+1)\pi}{b})^{2}-(\frac{(2m+1)\pi}{a_{1}})^{2}}\cdot
c}\nonumber\\&& \ \int^{b}_{0}e^{-ik\sin\alpha x_{0}}\sin
\frac{(2n+1)\pi x_{0}}{b}dx_{0} \int^{a}_{0}e^{-ik\sin\beta y_{0}}
\sin \frac{(2m+1)\pi y_{0}}{a}dy_{0}.
\end{eqnarray}

Equation (24) is the $C_{60}$ diffraction wave function of the
first slit, and the diffraction wave function $\psi_{2 P}$ for the
second slit can be obtained by making the coordinate translations
$x'=x, y'=y-(a+d), z'=z$, it is
\begin{eqnarray}
\psi_{2 P}(x,y,z,t)&=&-\frac{e^{ikR}}{4\pi
R}e^{-\frac{i}{\hbar}Et}\sum_{m=0}^{\infty}\sum_{n=0}^{\infty}\frac{16A}{(2m+1)(2n+1)\pi^2}
e^{i\sqrt{\frac{2ME}{\hbar^{2}}-(\frac{(2n+1)\pi}{b})^{2}-(\frac{(2m+1)\pi}{a})^{2}}\cdot
c}\nonumber\\&&
[i\sqrt{\frac{2ME}{\hbar^{2}}-(\frac{(2n+1)\pi}{b})^{2}-(\frac{(2m+1)\pi}{a})^{2}}+(ik-\frac{1}{R})
\sqrt{\cos^{2}\alpha-(\frac{s}{R})^{2}}] \nonumber\\&&
\int^{b}_{0}e^{-ik\sin\alpha\cdot
x^{'}}\sin\frac{(2n+1)\pi}{b}x^{'}dx^{'} \nonumber\\&&
\int^{2a+d}_{a+d}e^{-ik\sin\beta\cdot y^{'}} \sin
\frac{(2m+1)\pi}{a}(y^{'}-(a+d))dy^{'},
\end{eqnarray}
where $d$ is the two slit distance. The total diffraction wave
function for the double-slit is
\begin{eqnarray}
\psi_{P}(x,y,z,t)=c_{1}\psi_{1P}(x,y,z,t)+c_{2}\psi_{2P}(x,y,z,t),
\end{eqnarray}
where $c_{1}$ and $c_{2}$ are superposition coefficients , and
$|c_{1}|^{2}+|c_{2}|^{2}=1$. For the $C_{60}$ single-slit
diffraction, we can obtain the relative diffraction intensity $I$
on the display screen,

\begin{equation}
I\propto|\psi_{1P}(x,y,z,t)|^{2}.
\end{equation}

For the $C_{60}$ double-slit diffraction, we can obtain the
relative diffraction intensity $I$ on the display screen,
\begin{eqnarray}
I&\propto&|\psi_{P}(x,y,z,t)|^{2}\nonumber\\&=&
{c_{1}^{2}}|\psi_{1 P}(x,y,z,t)|^{2}+c_{2}^{2}|\psi_{2
P}(x,y,z,t)|^{2} +2c_{1}c_{2}Re[\psi_{1 P}^*(x,y,z,t)\psi_{2
P}(x,y,z,t)].
\end{eqnarray}
From Eqs. (27) and (28), we can obtain the relation between
diffraction intensity and diffraction angle. In Ref. [16], their
experiment data is about the relation between diffraction
intensity and diffraction position. The relation is
\begin{equation}
sin[\beta]=\frac{s}{R}=\frac{s}{\sqrt{l^2+s^2}}.
\end{equation}
Where $l$ is the distance from the slit to display screen, and $s$
is diffraction position.

\maketitle {\bf 4. Decoherence effect in double-slit
diffraction}\vskip 8pt

Decoherence is introduced here using a simple phenomenological
theoretical model that assumes an exponential damping of the
interferences [7, 20, 21], i.e., the decoherence is the dynamic
suppression of the interference terms owing to the interaction
between system and environment. The Eq. (26) describes the
coherence state coherence superposition, without considering the
interaction of system with external environment. When we consider
the effect of external environment, the total wave function of
system and environment for the double-slit factorizes as [7]
\begin{eqnarray}
\psi_{out}(x,y,z,t)=c_{1}\psi_{out_{1}}\otimes |E_{1}>_{t}
  +c_{2}\psi_{out_{2}}\otimes|E_{2}>_{t},
\end{eqnarray}
where $|E_{1}>_{t}$ and $|E_{2}>_{t}$ describe the state of the
environment. The diffraction intensity on the screen is now given
by[7]:
\begin{eqnarray}
I=(1+|\alpha_{t}|^{2})({c_{1}^{2}}|\psi_{out1}(x,y,z,t)|^{2}+c_{2}^{2}|\psi_{out2}(x,y,z,t)|^{2}
+2c_{1}c_{2}\Lambda_{t}Re[\psi_{out1}^*(x,y,z,t)\psi_{out2}(x,y,z,t)]).
\end{eqnarray}
where $\alpha_{t}=_{t}<E_{2}|E_{1}>_{t}$, and
$\Lambda_{t}=\frac{2|\alpha_{t}|}{1+|\alpha_{t}|^{2}}$. Thus,
$\Lambda_{t}$ is defined as the quantum coherence degree.  In Eq.
(31), the two slits wave functions $\psi_{1P}$ and $\psi_{2P}$ are
calculated by the quantum approach (in Eqs. (24)-(25)). In Refs.
[7], the two slits wave functions are two Gaussian wave packets.
The fringe visibility of $\nu$ is defined as [7]:
\begin{equation}
\nu=\frac{I_{max}-I_{min}}{I_{max}+I_{min}}.
\end{equation}
Where $I_{max}$ and $I_{min}$ are the intensities corresponding to
the central maximum and the first minimum next to it,
respectively. The value for the fringe visibility of $\nu=0.50$ is
obtained in Zeilinger et. al. experiment [18] ($I_{max}=880,
I_{min}=300$), and the quantum coherence degree
$\Lambda_{t}\approx\nu$ [7].

\vskip 8pt {\bf 5. Numerical result} \vskip 8pt

Next, we present our numerical calculation of relative diffraction
intensity. The main input parameters are: $C_{60}$ molecules mass
$M=1.4\times10^{-24}$kg, and Planck's constant
$\hbar=1.055\times10^{-34}$Js. For the single-slit experiment
[18], the $C_{60}$ velocity $v=220m/s$ (corresponding to $C_{60}$
wave length $\lambda=250\AA$), the slit width $a=10\mu m$, and the
distance between slit and display screen $l=2.29$m. In our
calculation, we take the same experiment parameters above, and the
theoretical input amplitude parameters are: $A=2.87\times
10^{14}$, the diffraction angle on $yz$ surface $\alpha=0$ rad,
the slit length $b=0.01m$ and the slit thickness $c=1.3\mu m$.
From Eq. (27), we can obtain the single slit diffraction intensity
pattern, and it is shown in FIG. 3. In FIG. 3, the solid curve is
our calculation result, and the dot curve is the experiment data
[18]. From FIG. 3, we can find the calculation result is agreement
with experiment data. For the double-slit diffraction, we consider
two cases: coherence superposition and decoherence effect. For
coherence superposition, from Eq. (28), we can calculate the
diffraction intensity and it is shown in FIG. 4. The experiment
parameters are: the $C_{60}$ wavelength $\lambda=250\AA$, the
first and second slit width, $a=0.05\mu m$, the distance between
the two slit $d=0.05\mu m$, the slit thickness $c=1.3\mu m$, the
distance between slit and display screen $l=1.25$m, In our
calculation, we take the same experiment parameters above, and the
theoretical input parameters are superposition coefficients
$c_{1}=0.566$, $c_{2}=0.824$ ($|c_{1}|^{2}+|c_{2}|^{2}=1$) and
amplitude parameter $A=1.27\times 10^{22}$. In FIG. 4, the solid
curve is our theoretical calculation, and the dot curve is the
experiment data [18]. From the FIG. 4, we find that the
theoretical result is in accordance with the experiment data, when
the position $s$ is in the range of $|s|\geq 100 \mu m $. When the
position $s$ is in the range of $|s|\leq 100 \mu m $, the
theoretical result has a large discrepancy with the experiment
data. We find the discrepancy can be eliminated when the
decoherence effect is considered. From Eq. (31), we can obtain the
diffraction intensity pattern and it is shown in FIG. 5. In
calculation, superposition coefficients $c_{1}=0.565$,
$c_{2}=0.824$, amplitude $A=1.69\times10^{22}$ and quantum
coherence degree $\Lambda_{t}=0.50$. From FIG. 5, we can find that
the new calculation result is more improvement than it in FIG. 4,
and it is in accordance with the experiment data, i.e., when the
decoherence effect is considered, the discrepancy between the
theoretical result and experiment data can be eliminated.

 \vskip 10pt

{\bf 6. Conclusion} \vskip 8pt In conclusion, we study $C_{60}$
single and double-slit diffraction with quantum theory approach.
The theoretical result of single-slit is in accordance with the
experiment data. For the double-slit diffraction, we study the
diffraction intensity by two approaches, which are the coherence
superposition and decoherence mechanism. When we only consider the
superposition of coherence, we find the theoretical result has a
large discrepancy with the experiment data, and when we consider
the decoherence mechanism, we find the theoretical result is in
accordance with the experiment data. So, the decoherence mechanism
is important for the double-slit diffraction of the large
molecules. We think the new quantum theory approach has universal
applicability, such as, it can study electron, atom and molecular
diffraction diffraction, and it can also be studied multi-slit and grating diffraction.  \\

\newpage
\begin{figure}[tbp]
\begin{picture}(50,35)
{\resizebox{12cm}{8cm}{\includegraphics{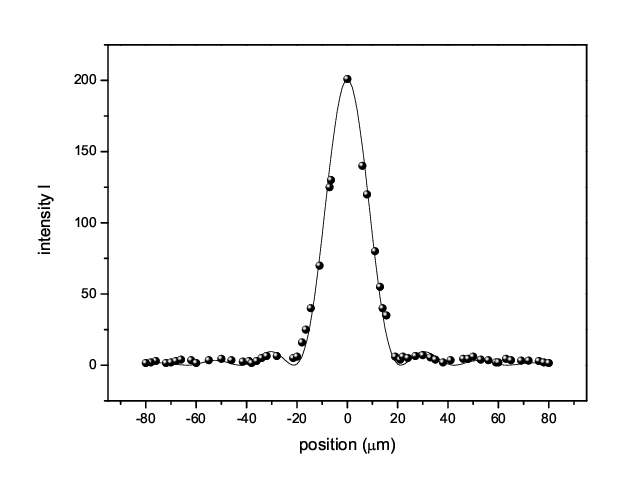}}}
\end{picture}
 \vskip 5pt
{\hspace{0.1in}FIG. 3: Comparison between theoretical prediction
from Eq. (27) (solid line) and \\ experimental data taken from
[18](circle point) for $C_{60}$ single-slit diffraction.}
 \label{moment}
\end{figure}
\begin{figure}[tbp]
\begin{picture}(50,35)
{\resizebox{12cm}{8cm}{\includegraphics{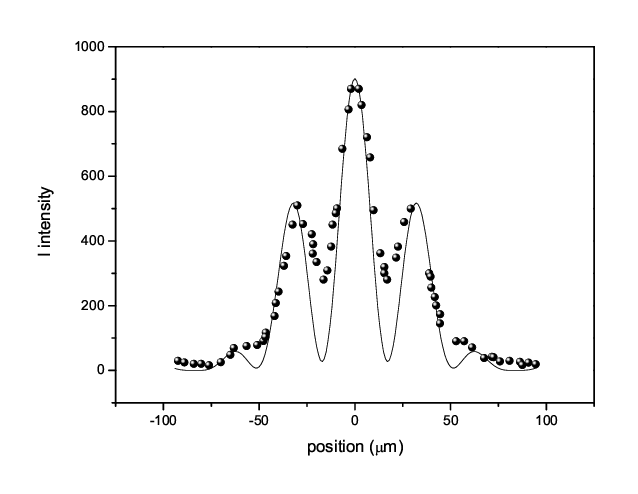}}}
\end{picture}
 \vskip 5pt
{\hspace{0.1in}FIG. 4: Comparison between theoretical prediction
from Eq. (28) (solid line) and experimental data taken \\ from
[18](circle point) for $C_{60}$ double-slit diffraction, no
including the decoherence effects}
 \label{moment}
\end{figure}

\begin{figure}[tbp]
\begin{picture}(50,40)
{\resizebox{12cm}{8cm}{\includegraphics{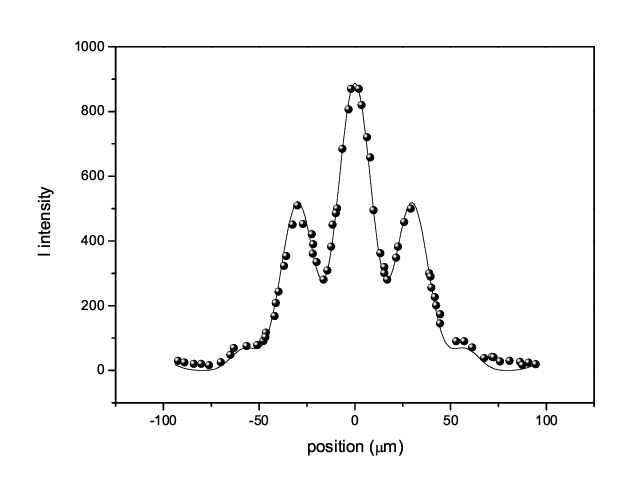}}}
\end{picture}
 \vskip 5pt
{\hspace{0.1in}FIG. 5: Comparison between theoretical prediction
from Eq. (31) (solid line) and experimental data taken \\ from
[18](circle point) for $C_{60}$ double-slit diffraction, including
the decoherence effects.} \label{moment}
\end{figure}

\end{document}